# Chapter 2

# What constitutes a nanoswitch? A Perspective

\* ***Supriyo Datta and Vinh Quang Diep,*** Purdue University
\*\* ***Behtash Behin-Aein,*** GLOBALFOUNDRIES Inc.

## Contents

\* [datta@purdue.edu](mailto:datta@purdue.edu)        \*\* [Behtash.Behin-Aein@globalfoundries.com](mailto:Behtash.Behin-Aein@globalfoundries.com)

# ABSTRACT

Progress in the last two decades has effectively integrated spintronics and nanomagnetics into a single field, creating a new class of spin-based devices that are now being used both to *Read (R)* information from magnets and to *Write (W)* information onto magnets. Many other new phenomena are being investigated for nanoelectronic memory as described in Part II of this book. It seems natural to ask whether these advances in memory devices could also translate into a new class of logic devices.

What makes logic devices different from memory is the need for one device to drive another and this calls for gain, directionality and input-output isolation as exemplified by the transistor. With this in mind we will try to present our perspective on how *W* and *R* devices in general, spintronic or otherwise, could be integrated into transistor-like switches that can be interconnected to build complex circuits without external amplifiers or clocks.

We start with a very brief and oversimplified discussion of the most common switch used to implement digital logic based on complementary metal oxide semiconductor (CMOS) transistors. We will argue that a CMOS switch can be viewed as an integrated *W-R* unit having an input-output asymmetry that give it gain and directionality. Such a viewpoint uses the word "Write" in an unconventional sense, and is not intended to provide any insight into the operation of CMOS switches, but rather as an aid to understanding how *W* and *R* units based on spins and magnets can be combined to build transistor-like switches.

Next we will discuss the standard *W* and *R* units used for magnetic memory devices and present one way to integrate them into a single unit with the input electrically isolated from the output. But we argue that this integrated *W-R* unit would not provide the key property of gain. We will then show that the recently discovered giant spin Hall effect (GSHE) could be used to construct a *W-R* unit with gain and suggest other possibilities for spin switches with gain.

We end with a brief evaluation of these alternative switches in terms of possible applications. A key metric is the energy-delay product and it appears that new materials and phenomena for *W* and *R* units will be needed to provide any improvement over standard CMOS switches. On the other hand the non-volatility and reconfigurability of switches based on magnets is a novel feature that could enable a new class of circuits very different from those currently possible.

\* [datta@purdue.edu](datta@purdue.edu)               \*\* [Behtash.Behin-Aein@globalfoundries.com](Behtash.Behin-Aein@globalfoundries.com)

## 2.1 The search for a better switch

A basic element in digital logic is a switch or an inverter comprising a pair of complementary metal oxide semiconductor (CMOS) nanotransistors (Fig.2.1.1a) whose resistances $R_1$ and $R_2$ change in a complementary manner in response to the input voltage $V_{in}$. As $V_{in}$ changes from 0 to $V_{DD}$, the resistance $R_1$ of the "NMOS" transistor gets smaller while the resistance $R_2$ of the "PMOS" transistor gets larger making the output voltage

$$V_{out} = V_{DD} \frac{R_1}{R_1 + R_2}$$

change from $V_{DD}$ to 0 as shown in Fig.2.1.1b so that the output represents an inverted version of the input.

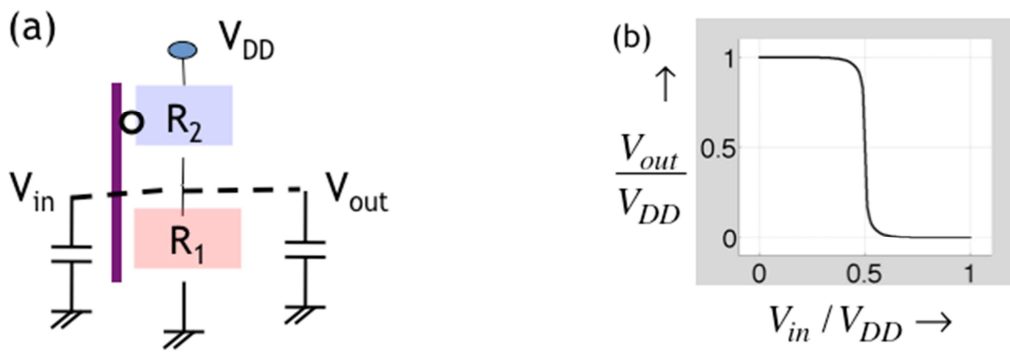

*Fig.2.1.1 (a) CMOS inverter comprises an NMOS transistor ($R_1$) and a PMOS transistor ($R_2$). (b) Input-output characteristics of the inverter.*

For some time now it has been recognized that one of the biggest obstacles to continued downscaling is the heat dissipated [1][2]. Every time a switch changes state, the charge $Q$ stored in an input or an output capacitors gets dumped thus dissipating an energy of $QV_{DD}$. If there are $N_{act}$ number of active switches switching at a frequency $f$ per second, the power dissipated can be written as

$$P = N_{act} Q V_{DD} f \qquad (2.1.1)$$

* datta@purdue.edu                     ** Behtash.Behin-Aein@globalfoundries.com

To estimate the energy QV$_{DD}$ dissipated per switch, we could use the numbers for the Intel® Core™ i3-530 Processor taken from their website at http://ark.intel.com/products/46472 [3]

$$P = 73\,Watts\,,\ f = 2.93\,GHz$$

$$N_{act} = \underbrace{559 \times 10^6}_{\text{Total number of transistors}} \times \underbrace{10\%}_{\text{activity factor}}$$

$$QV_{DD} \approx \frac{73W}{2.93GHz \times (559 \times 10^6 \times 10\%)} = 2785\,eV \approx 445\,aJ$$

Since the power dissipated cannot increase too much beyond 73W we cannot increase the number of active switches $N_{act}$ or their speed of operation $f$ very much, unless we discover switches that dissipate less energy without compromising the speed. It was this recognition that prompted the Semiconductor Research Corporation (SRC) together with the National Science Foundation (NSF) to launch the Nanoelectronics Research Initiative (NRI) back in 2005 with the objective of exploring the possibility of realizing a better switch based on any known physical mechanism.

*Outline:* In this chapter we would like to share our perspective on the question of what constitutes a "transistor-like switch", and how we could build one based on novel physical mechanisms and assess its performance. As an example of a radically different physical mechanism, we will focus on spintronics and nanomagnetics where there has been enormous progress in the last two decades. But we will try to phrase our discussion and conclusions in general terms so that it could be easily adapted to other phenomena as well.

The new discoveries in spins and magnets are already finding use in memory devices both to *Read(R)* information from magnets and to *Write(W)* information onto magnets. Many other new phenomena are being investigated for nanoelectronic memory as described in Part II of this book. It seems natural to ask whether these advances in W&R units for memory devices could also translate into a new class of logic devices.

In *Section 2.2* we start with a very brief and oversimplified discussion of the most common switch used to implement digital logic based on CMOS transistors stressing the key property of gain that allows us to interconnect them into complex circuits without the use of external amplifiers or clocks. To harness spins and magnets for logic applications one could either integrate them onto CMOS devices that provide the gain (see for example [4]) or try to design transistor-like spin-magnet devices having gain. It is the *latter* option that we will explore in this chapter.

* datta@purdue.edu   ** Behtash.Behin-Aein@globalfoundries.com

We will argue that a CMOS switch can be viewed as an integrated *W-R* unit, using the word "Write" in a somewhat unconventional sense. The purpose is not to provide any new insight into CMOS, but to help understand how *W* and *R* units used for memory devices can be combined to build transistor-like switches.

In *Section 2.3* we discuss the standard *W* and *R* devices used for magnetic memory devices and present one way to integrate them into a single unit where the input and output are electrically isolated, but we argue that such a unit would not provide the key transistor-like property of gain. We will then show (*Section 2.4*) that the recently discovered giant spin Hall effect (GSHE) could be used to construct a *W-R* unit with gain[5].

Other possibilities for transistor-like W-R units with gain are briefly discussed in *Section 2.5* including all-spin logic (ASL)[6] along with new possibilities based on newly discovered phenomena. Indeed, with the growing research interest in STT-MRAM (spin transfer torque magnetic random access memory) for both stand-alone[7] and embedded memory applications[8][9] it is likely that many more new phenomena will be discovered that could be used to construct transistor-like W-R units for logic applications.

Also we should mention that there are other independent proposals like the trans-spinor [10] and m-logic[11] that could be viewed as examples of the same W-R paradigm for logic that we are discussing here.

In *Section 2.6* we end with a brief discussion of how these alternative transistor-like switches could be evaluated in terms of possible applications. A key metric is the energy-delay product and we will argue that new materials and phenomena for *W* and *R* units are needed to provide any improvement over standard CMOS switches. On the other hand the non-volatility and reconfigurability of switches based on magnets is a novel feature that could enable a whole new class of circuits very different from those currently possible.

* datta@purdue.edu             ** Behtash.Behin-Aein@globalfoundries.com

### 2.2 Complementary metal oxide semiconductor switch: Why it shows gain

To understand the key characteristics of a transistor-like switch it is useful to take a brief look at a standard transistor. The simplest transistor is an NMOS or a PMOS, but we choose a CMOS switch which combines the two into a single switch that performs a logic operation, namely NOT, and has an input-output characteristic resembling those obtained from the spin switches discussed later in the chapter.

A CMOS switch is made of an NMOS and a PMOS transistor, which constitute the voltage controlled resistors $R_1$ and $R_2$ shown in Fig.2.1.1a. Let us briefly describe the characteristics of an NMOS and a PMOS transistor, which can then be combined to obtain the input-output characteristics of the CMOS inverter shown in Fig.2.1.1b.

*NMOS transistor:* The resistor $R_1$ in Fig.2.1.1a is an NMOS transistor whose resistance

$$R_1 = V_{out} / I_1$$

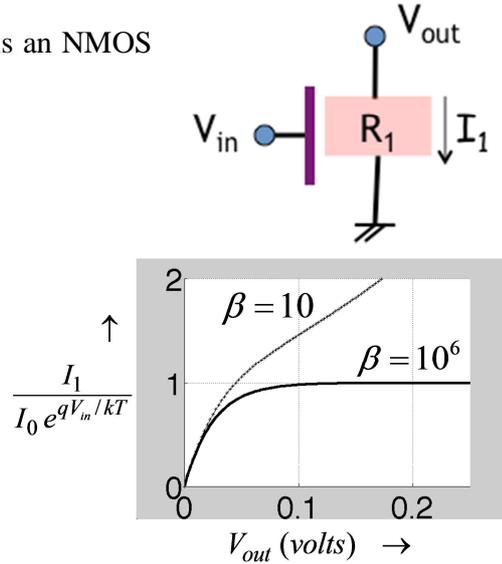

is reduced by a positive input voltage $V_{in}$. For small input voltages the conductance ($1/R_1$) increases exponentially with $V_{in}$ (see for example [12])

$$\frac{I_1}{V_{out}} \sim e^{qV_{in}/kT}$$

Also, the resistance is not constant and ideally the current saturates for large $V_{out}$. We could describe this behavior approximately as ($I_0$: constant)

$$I_1 = I_0 e^{qV_{in}/kT} \times e^{qV_{out}/\beta kT} \times \left(1 - e^{-qV_{out}/kT}\right) \qquad (2.2.1)$$

With $\beta = 10^6$ the current saturates perfectly which is what we would ideally like, but with $\beta = 10$ we have a characteristic looking more like real transistors, with the current showing an increase with $V_{out}$ due to "drain-induced barrier lowering (DIBL)."

---

* *datta@purdue.edu*                    ** *Behtash.Behin-Aein@globalfoundries.com*

**PMOS transistor:** The other resistor ($R_2$) in Fig.2.1.1a is a PMOS transistor whose resistance is increased by a positive input voltage and we will assume that the characteristics can be described by an expression similar to Eq.(2.2.1) but with $V_{in}$ and $V_{out}$ replaced by $(V_{DD} - V_{in})$ and $(V_{DD} - V_{out})$ respectively.

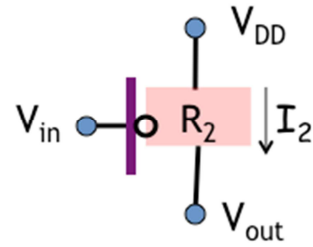

$$I_2 = I_0 e^{q(V_{DD}-V_{in})/kT} \times e^{q(V_{DD}-V_{out})/\beta kT} \times \left(1 - e^{-q(V_{DD}-V_{out})/kT}\right) \quad (2.2.2)$$

In general the NMOS and PMOS need not be symmetric with the same constant $I_0$ appearing in both current expressions (see Eqs.(2.2.1) and (2.2.2)) but we will ignore such "details", since our objective is to use the simplest model just to illustrate the main points.

**Switch characteristics:** The input-output characteristics of a CMOS inverter are obtained by solving Eqs.(2.2.1) for $R_1$ (NMOS) and (2.2.2) for $R_2$ (PMOS) simultaneously. For any particular $V_{in}$, we adjust $V_{out}$ numerically so as to make $I_1 = I_2$. This leads to the switch characteristics shown in Fig.2.2.1 for different values of the parameter $\beta$ reflecting different degrees of current saturation as discussed earlier.

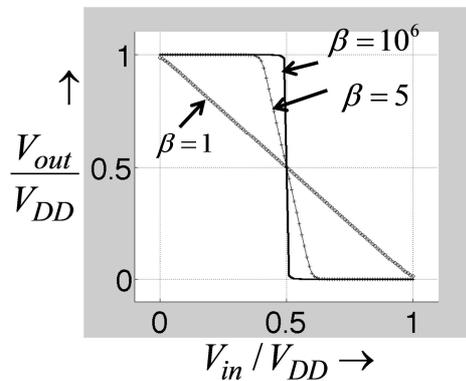

*Fig.2.2.1 Input-output characteristics of a CMOS inverter obtained by solving Eqs.(2.2.1) for $R_1$ (NMOS) and (2.2.2) for $R_2$ (PMOS) simultaneously for different values of $\beta$.*

**Gain:** A key attribute of a logic unit is its **gain** defined as the change in the output voltage for a given change in the input voltage

$$Gain \equiv \frac{\Delta V_{out}}{\Delta V_{in}} \quad (2.2.3)$$

* datta@purdue.edu          ** Behtash.Behin-Aein@globalfoundries.com

A logic unit should have a gain > 1 in order to drive another in a circuit. It is evident from Fig.2.2.1 that while with large values of $\beta$ the inverter has a sizeable gain, the gain ≤ 1 if $\beta = 1$. Our $\beta$ is a parameter introduced (see Eqs.(2.2.1), (2.2.2)) to account for the drain voltage dependence of the current and is usually far bigger than one for any real transistor. And so the situation with $\beta = 1$ is not of any real practical significance.

We are simply using the factor $\beta$ to make the point that in order to have gain>1, one needs an input-output asymmetry whereby the current is controlled largely by $V_{in}$, and very little by $V_{out}$. This is evident if we rewrite the current in Eq.(2.2.1) for large $V_{out}$

$$I_1 \approx I_0 \, e^{qV_{in}/kT} \times e^{qV_{out}/\beta kT}$$

showing that the factor $\beta$ represents the ***asymmetry*** in the response of the current to the input and output voltages. With $\beta = 1$, this asymmetry is lost and so is the gain.

***Switch as a Write-Read pair:*** Before we move on, let us point out that a CMOS switch could be viewed as a *Write(W) – Read(R)* pair, if we use the word Write in a somewhat unconventional sense. This viewpoint may seem artificial and probably does not provide any insight into the operation of CMOS switches. Our reason for introducing it is as an aid to understand how Write and Read units based on spins and magnets can be combined to form switches.

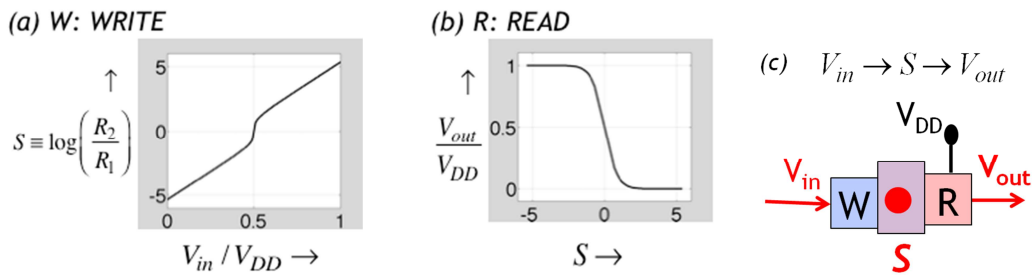

*Fig.2.2.2 (a) "Write" operation in a CMOS inverter: State of the CMOS pair defined as $S = \log(R_2 / R_1)$ as a function of the input voltage $V_{in}$. (b) "Read" operation : Output voltage $V_{out}$ as a function of the state S. (c) Symbolic representation depicting the switch as a Write-Read pair.*

\* <u>datta@purdue.edu</u>        \*\* <u>Behtash.Behin-Aein@globalfoundries.com</u>

We could define the state $S$ of the complementary pair in terms of the ratio of the two resistances $R_1$ and $R_2$:

$$S = \log\left(\frac{R_2}{R_1}\right) \qquad (2.2.4)$$

and create two plots: the *Write(W)* characteristics showing $S$ as a function of the input voltage $V_{in}$ (Fig.2.2.2a) and the *Read(R)* characteristics showing the output voltage $V_{out}$ as a function of the state $S$ (Fig.2.2.2b).

The former describes the $W$ operation whereby the state $S$ of the CMOS inverter is set according to the input voltage $V_{in}$, while the latter describes the $R$ operation in that the supply voltage $V_{DD}$ results in an output voltage $V_{out}$ depending on the state $S$ as shown symbolically in Fig.2.2.2c:

$$V_{in} \rightarrow S \rightarrow V_{out} \qquad (2.2.5)$$

Note that we are stretching the meaning of the *Write* operation somewhat, since the state S does not persist once the input $V_{in}$ has been removed: Unlike real memory devices, the Read operation needs to be carried out while $V_{in}$ is present. Our purpose here is simply to connect the language of memory devices involving $W$ and $R$ units to that of CMOS so that we can understand and adapt the key property of gain that distinguishes logic units.

To integrate $W$ and $R$ into a transistor-like switch, an input-output asymmetry seems important: the gain of a CMOS switch seems intimately related to the fact that the input voltage $V_{in}$ is far more effective in controlling the state of the switch $S$ than the output voltage $V_{out}$. This input-output asymmetry and the resulting gain make a transistor very different from reversible Hamiltonian systems often discussed in the context of nanoscale systems. To harness spins and magnets for logic devices we have two broad options:
- integrate them onto CMOS devices which provide the gain, or
- design transistor-like spin-magnet devices that have gain.

It is the latter possibility that we are discussing in this chapter.

* datta@purdue.edu ** Behtash.Behin-Aein@globalfoundries.com

**2.3 A switch based on magnetic tunnel junctions: Would it show gain?**

Since *W* and *R* units based on magnetic tunnel junctions (MTJ's) are now well-known, it seems natural to ask whether these could be combined into a transistor-like switch. Before addressing that question let us briefly summarize how an MTJ-based *W* and *R* device works.

***Operation of a magnetic tunnel junction (MTJ):*** Fig.2.3.1a shows a simplified magnetic tunnel junction (MTJ) structure having one layer with a reference magnet $\hat{M}$ separated by a tunneling barrier from a free layer magnet of nanometer scale thickness whose magnetization $\hat{m}$ represents the stored information. Fig.2.3.1b shows a typical resistance versus current characteristic of an MTJ taken from Kubota et al. [13] which illustrates the basic physical phenomena underlying both the *R* and *W* operations.

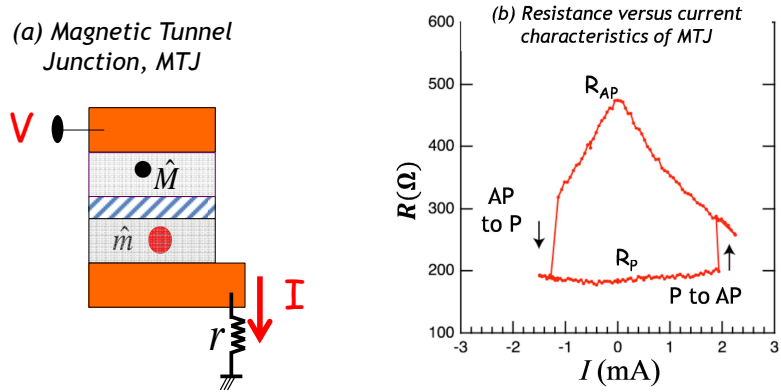

*Fig.2.3.1 (a) A simplified schematic of a magnetic tunnel junction (MTJ) (b) A typical resistance versus current characteristic of an MTJ taken from Kubota et al. Reprinted by permission from Macmillan Publishers Ltd: Nature Physics, Ref. [13], copyright 2008.*

At low currents the resistance can have one of two values: The smaller one ($R_P$) corresponds to the ***P*** configuration with the two magnetizations ***parallel***: $\hat{m} \parallel +\hat{M}$ while the larger one corresponds to the ***AP*** configuration with the magnetizations ***anti-parallel***: $-\hat{m}\parallel+\hat{M}$. This phenomenon allows one to *Read* the state of the free magnet $\hat{m}$ relative to the fixed magnet $\hat{M}$ by applying a small voltage $V$.

On the other hand Fig.2.3.1b shows that at sufficiently high positive currents the free layer switches from a ***P*** to an ***AP*** configuration while at high negative currents it switches from an ***AP*** to a ***P*** configuration. This phenomenon allows one to *Write* information contained in the polarity of the current onto the magnetization of the free layer.

\* [datta@purdue.edu](datta@purdue.edu)          \*\* [Behtash.Behin-Aein@globalfoundries.com](Behtash.Behin-Aein@globalfoundries.com)

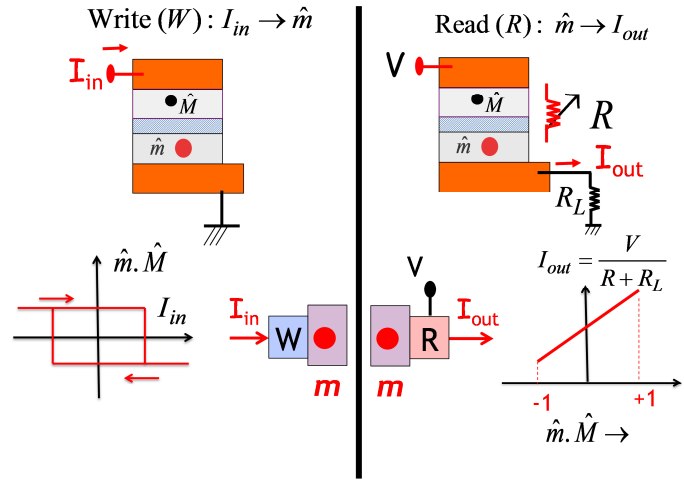

*Fig.2.3.2 An MTJ unit can be used either as a Write (W) unit or as Read (R) unit.*

Fig.2.3.2 shows the basic characteristics of the Write and Read unit based on MTJ device and their symbolic operations. The Write unit converts the input current $I_{in}$ into the magnetization $\hat{m}$ of the free magnet, while the Read unit converts the information stored in $\hat{m}$ into an output current $I_{out}$ given by

$$I_{out}(\hat{m}) = \frac{V}{R(\hat{m}) + R_L} \tag{2.3.1}$$

where V is the supply voltage and $R_L$ is a fixed load resistance.

*A W-R unit with electrical isolation:* We can now proceed to combine an MTJ *Read (R)* device with a *Write (W)* device to obtain a composite unit as shown in Fig.2.3.3a where the magnet $\hat{m}$ from *R is* coupled to the magnet $\hat{m}'$ from *W* through their dipolar magnetic field as indicated by a dashed line. This allows the information to propagate from input to output: An input current $I_{in}$ switches the *Write* magnet ($\hat{m}'$) which in turn switches the *Read* magnet ($\hat{m}$) through the dipolar coupling causing a change in the output current $I_{out}$ as described by Eq.(2.3.1). At the same time the input is electrically isolated from the output allowing these units to be interconnected to form large circuits. This feature has some similarity to m-logic[11] which proposes to use exchange coupling to couple input domains to the output.

The W-R unit in Fig. 2.3.3a can be modeled with an equivalent circuit of the form shown in Fig.2.3.3c which leads to the overall input-output characteristic shown in Fig. 2.3.3d.

* *datta@purdue.edu*                  ** *Behtash.Behin-Aein@globalfoundries.com*

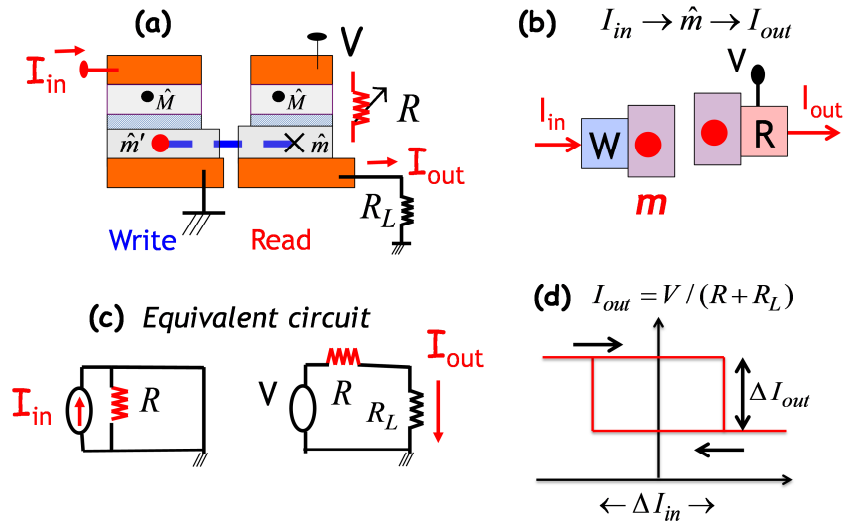

*Fig.2.3.3 (a) A Write (W) and a Read (R) unit combined to obtain a logic unit with input-output isolation. The dashed line represents the magnetic coupling between the two free magnets. (b) Representative symbol (c) Equivalent circuit (d) Input-output characteristic.*

***Does this W-R unit have gain?*** So far things seem straightforward, combining a *W* unit with an *R* unit using magnetic coupling to allow information transfer while maintaining electrical isolation. But can this unit exhibit gain, so that the swing in the output current will exceed that in the input current?

The swing in the output current is proportional to the voltage as in Eq. (2.3.1) or Fig.2.3.3d. It seems that we could make it exceed $\Delta I_{in}$ simply by choosing a large enough supply voltage $V$.

The problem, however, is this. The voltages $V$ of MTJ's in the *Read* unit also give rise to a spin current $I_s$ that acts on the *Read* magnet $\hat{m}$. Ordinarily *Read* voltages are kept small enough such that the resulting spin current $I_s$ does not disturb the free layer whose information we are trying to read. But if we do that, the output current would be much smaller than what is needed as input to drive the next stage and we could not build circuits without using an external amplifier.

To obtain an output spin current comparable to the critical spin current needed to drive the input of the next stage we have to make the supply voltage $V$ even larger and the resulting spin current $\vec{I}_S$ acting on the *Read* magnet $\hat{m}$ would exceed the spin current $\vec{I}_S{'}$ acting on the *Write* magnet $\hat{m}'$. The state of the magnet $\hat{m}$ will then be determined by the output rather than the input. This is analogous to building a CMOS inverter (section 2.2) using transistors whose current is controlled more strongly by the drain than by the gate, described by a $\beta$ lesser than one.

* *datta@purdue.edu*                    ** *Behtash.Behin-Aein@globalfoundries.com*

We need to design the magnet pair such that it "feels" the influence of the input current far more strongly than that of the output current. It may be possible to design composite magnets with different materials to achieve this, but a relatively straightforward design seems possible utilizing a relatively recent discovery, namely the giant spin Hall effect (GSHE) as we will describe next.

## 2.4 The giant spin Hall effect: A route to gain

The spin Hall effect is exhibited by materials with spin-orbit coupling where the flow of current $I$ is accompanied by a spin current $I_s$ in the perpendicular direction, such that the *spin current density* equals the *charge current density* times the spin Hall angle [14]:

$$\frac{I_S}{L} = \theta_H \frac{I}{t}$$

so that 
$$I_S = \underbrace{\theta_H \frac{L}{t}}_{\equiv \beta} I \quad (2.4.1)$$

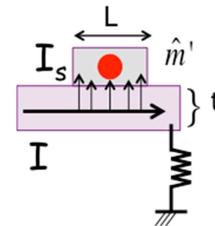

The spin Hall angle $\theta_H$ is usually quite small, but recently a number of GSHE materials have been discovered which have $\theta_H$ values as high as 0.3 [15]. More interestingly, a proper choice of geometry with L >> t can give values of $\beta$ in excess of one, corresponding to a spin current $I_s$ that exceeds the current $I$.

We can make use of this natural gain provided by the GSHE material, by replacing the MTJ based Write unit in Fig. 2.3.3a with the one shown in Fig.2.4.1a. To get better magnetic coupling between the *Write* and *Read* magnets it may be advisable to stack them vertically as shown in Fig.2.4.1b. In any case the W-R unit can be modeled with an equivalent circuit of the form shown in Fig.2.4.1c obtained by combining the *W* unit with a separate equivalent circuit for the *R* unit. Here $G_m$ is the conductance of the MTJ device and it can be related to $G$ and $\Delta G$ representing the sum and difference respectively of the parallel and anti-parallel conductances.

$$G_m = \frac{G}{2} + \frac{\Delta G}{2} \hat{m}.\hat{M} \quad (2.4.2)$$

* datta@purdue.edu  ** Behtash.Behin-Aein@globalfoundries.com

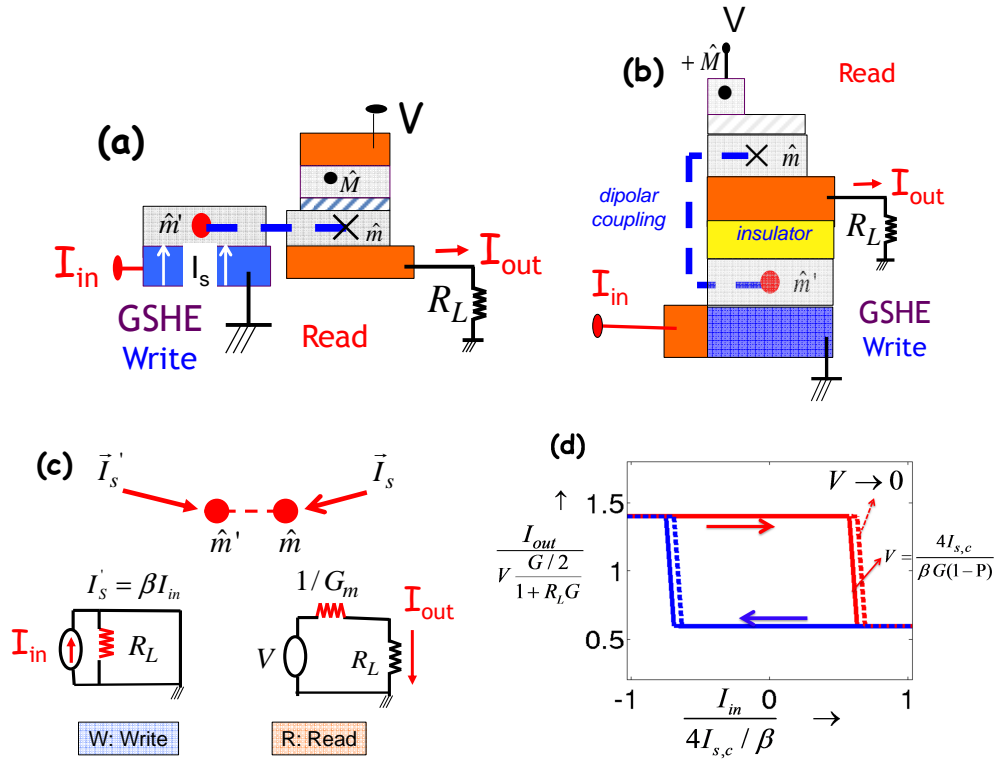

*Fig.2.4.1 (a) A GSHE Write and a Read unit combined to build a logic unit with input-output isolation and gain. The dashed line represents the magnetic coupling between the two free magnets. (b) Better magnetic coupling between the Write and Read magnets can be obtained by stacking the units vertically rather than laterally. (c) Equivalent circuit for structure in (a). (d) In-output characteristic.*

The parameters $G$ and $\Delta G$ can be related to experimentally reported quantities like the tunneling magnetoresistance (TMR)

$$TMR \equiv \frac{G_P}{G_{AP}} - 1 \qquad (2.4.3a)$$

or the polarization $P \equiv \Delta G / G$

$$P \equiv \frac{G_P - G_{AP}}{G_P + G_{AP}} = \frac{TMR}{TMR+2} \qquad (2.4.3b)$$

The input-output characteristic (Fig.2.4.1d) was obtained from the equivalent circuit in Fig.2.4.1c using the same method as described in [5] with the spin currents coupled to the Landau-Lifshitz-Gilbert (LLG) equation for the magnet pair $\hat{m}' - \hat{m}$. For a small supply voltage V (→ 0) the characteristic is symmetric about the origin but it shifts to the left as V is increased because the spin current $I_s$ injected by the *Read* unit makes it easier to switch from +1 to -1 than to switch from -1 to +1. Indeed if the voltage $V$ were too large we would not have a useful switch. But the GSHE allows us to use a relatively small $V$ and still have gain.

* datta@purdue.edu  ** Behtash.Behin-Aein@globalfoundries.com

The key point is that the gain $\beta$ from the GSHE allows the use of a relatively low voltage. A *Write* unit based on an ordinary spin-torque device (like the one discussed in the last Section) has a spin current that is less than the charge current $I$, corresponding to a $\beta$ less than one and thus requires a much larger voltage $V$ to drive the next unit.

***Concatenability:*** Although the switch in Fig.2.4.1 exhibits gain and input-output isolation, it is not "concatenable" because the output from the *Read* unit is purely positive (assuming V is positive) and is not appropriate for driving the *Write* unit of the next stage which requires a bipolar input that takes on both positive and negative values. One can think of two broad approaches to addressing this concatenability issues:
- design a *Write* unit that can be driven with purely positive voltages, and
- design a *Read* unit that produces a bipolar output.

One possible design, based on the second approach [5], is shown in Fig.2.4.2a. It requires a more complicated fabrication process since two MTJ's are required: If they had the same resistance the output voltage would be zero since one is connected to +V and one to –V. But the two MTJ's will never have the same resistance, since their fixed magnets are antiparallel, namely $+\hat{M}$ and $-\hat{M}$. Depending on whether the free layer magnetization $\hat{m}$ is parallel to $+\hat{M}$ or $-\hat{M}$, one MTJ will be in its low resistance or ***P*** configuration while the other will be in its high resistance or ***AP*** configuration.

If the MTJ connected to +V is in its low resistance state then the output will be closer to +V and hence positive. If the MTJ connected to –V is in its low resistance state then the output will be closer to –V and hence negative. This dual MTJ Read unit should thus do what we want, namely convert positive or negative magnetization into a bipolar (that is, positive or negative) output voltage. And hence a bipolar output current.

* *datta@purdue.edu*  ** *Behtash.Behin-Aein@globalfoundries.com*

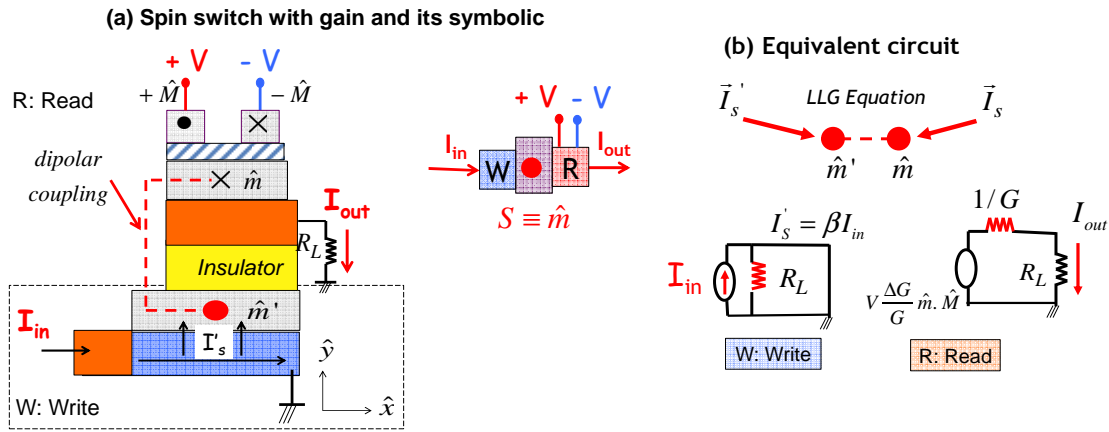

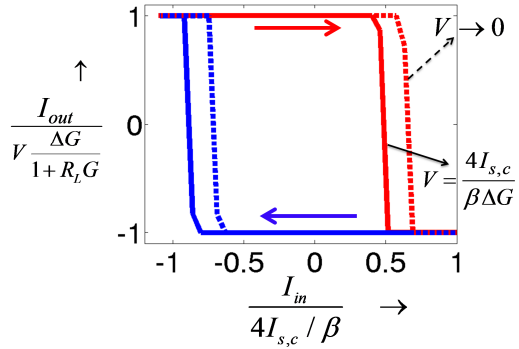

*Fig.2.4.2 (a) An integrated W-R unit obtained by vertically stacking a W device based on the GSHE and a dual MTJ R device with the corresponding magnets $\hat{m}'$ and $\hat{m}$ magnetically coupled. (b) Equivalent circuit for structure in (a). (c) Input-output characteristics of the device which also implies gain and non-volatility properties of the switch. Reprinted with permission from[5].Copyright 2012, AIP Publishing LLC.*

For quantitative modeling we could use the equivalent circuit shown in Fig.2.4.2b, where $G$ and $\Delta G$ are the same as the ones used in Eq. (2.4.2). This equivalent circuit shows *that the open circuit voltage is proportional to $\hat{m}\cdot\hat{M}$, the component of $\hat{m}$ along $\hat{M}$* giving an output current of

$$I_{out} = \frac{V\Delta G/G}{R_L + 1/G}\hat{m}\cdot\hat{M}$$

Fig.2.4.2c shows the input-output characteristic calculated using the equivalent circuit shown in Fig.2.4.2b and coupling the spin currents to the LLG equation for the magnet pair $\hat{m}'-\hat{m}$ [5].

* datta@purdue.edu             ** Behtash.Behin-Aein@globalfoundries.com

The gain can be estimated by noting that from Fig.2.4.2c

$$\frac{\Delta I_{out}}{V\dfrac{\Delta G}{1+R_L G}} \approx 2\frac{\Delta I_{in}}{4I_{s,c}/\beta}$$

so that 
$$Gain \equiv \frac{\Delta I_{out}}{\Delta I_{in}} \approx \frac{V\Delta G}{1+R_L G}\frac{\beta}{2I_{sc}} \qquad (2.4.4)$$

For an approximate gain of ~2, we need a voltage of

$$V \approx \frac{4I_{sc}}{\beta \Delta G}, \quad if \quad R_L G << 1$$

which only has a minor effect on the input-output characteristic (Fig.2.4.2c).

***Proof of gain and directionality:*** A good test for switches with gain and directionality is the following. It should be possible to connect an odd number of such switches to form a ring oscillator (Fig.2.4.3). If the voltages *V* on the *Read* units exceed the threshold value needed to drive the following *Write* unit, then each unit switches the next unit anti-parallel to itself. With an odd number of magnets, three in this case, in the loop, there is no way for all three units to be anti-parallel to each other and there is no satisfactory steady state. Unit 1 switches unit 2, which in turn switches unit 3, which goes on to switch unit 1 and the result is an oscillatory output as obtained from detailed simulation [5].

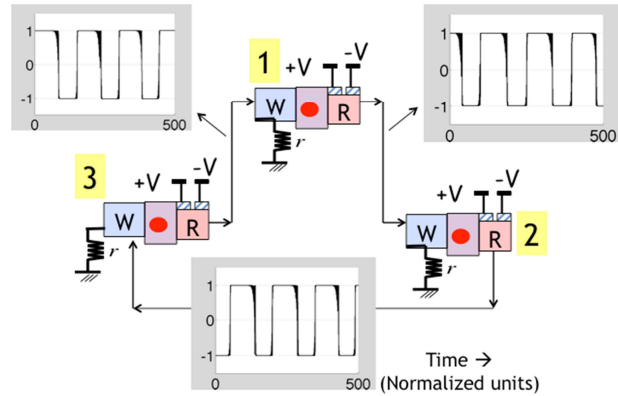

*Fig.2.4.3 An odd number of W-R units with gain and directionality can be connected in a ring to form a ring oscillator. Reprinted with permission from [5]. Copyright 2012, AIP Publishing LLC.*

Such oscillations are well known using an odd number of CMOS switches, and should provide a good test for the properties of gain and directionality which ensure that a signal can propagate without losing strength or compounding errors.

* *datta@purdue.edu*      ** *Behtash.Behin-Aein@globalfoundries.com*

## 2.5 Other possibilities for switches with gain

So far we have seen two examples of switches, the standard CMOS switch and a proposed one based on spins and magnets. Both can be viewed as integrated W-R units where the input $V_{in}$ or $I_{in}$ *write*s the internal state *S* which is then *read* to generate an output. As noted earlier, in the case of CMOS we are stretching the meaning of *Write* somewhat.

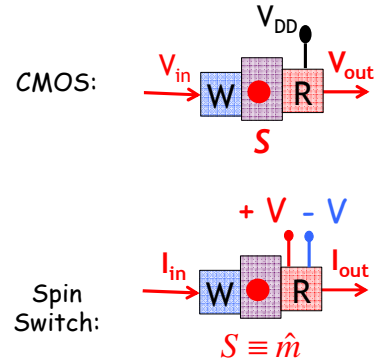

For the CMOS switch, the internal state *S* can be defined in terms of the resistances of the NMOS and PMOS transistors, while for the spin switch *S* represents the magnetizations of the magnet pair:

$$CMOS\ Switch: \quad S \equiv \log(R_2/R_1)$$
$$Spin\ Switch: \quad S \equiv \hat{m}$$

We have argued that a spin switch with gain could be implemented by combining a dual MTJ-based *Read* unit with a GSHE-based *Write* unit. However, this is by no means the only possibility. For example, the *Write* unit could involve a voltage controlled multiferroic[16] as shown in Fig.2.5.1

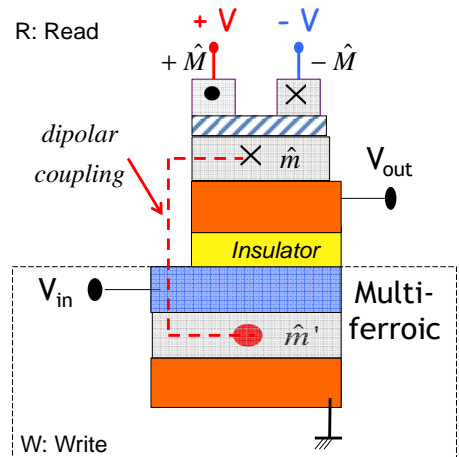

*Fig.2.5.1 Voltage controlled spin switch*

Note, however, that in order for one unit to be able to drive another, the output voltage $V_{out}$ has to be large enough to switch the multiferroic, thus requiring a minimum voltage *V* on the *Read* device. This voltage should not be too large, or the resulting spin current could control the magnet pair $\hat{m}'-\hat{m}$ instead of the input voltage. As we have argued, a suitable degree of input-output asymmetry is needed and whether it can be achieved has to be assessed carefully for each individual proposal. Indeed many other phenomena like voltage controlled magnetic anisotropy (VCMA) (see for example, [17-19]) could potentially be used to design improved W units.

* datta@purdue.edu                      ** Behtash.Behin-Aein@globalfoundries.com

A key difference with the CMOS switch is that unlike the resistance of an NMOS or a PMOS transistor, the magnetization represents a non-volatile state $S$. However, it may be possible to use other mechanisms for voltage-controlled resistance based on phase transition phenomena (like the Mott transition) [20] which could provide a non-volatile internal state $S$ like magnets.

***All-spin logic:*** Both the CMOS and the spin switch that we have described involve ordinary voltages and currents as the input and output variables and can be interconnected with ordinary wires to form circuits.

One could also envision switches based on *spin voltages* and *spin currents* as the input and output variables. An input spin current switches the *Write* magnet $\hat{m}'$ which through the dipolar coupling switches the *Read* magnet $\hat{m}$ causing the output spin current to switch (Fig.2.5.2). This is similar to the all-spin logic (ASL) device [6], [21] with the difference that the magnets $\hat{m}'$ and $\hat{m}$ are electrically isolated in the present version. It was shown theoretically in [6], [21] that switches with gain and directionality can be implemented with voltages as low as 10 mV.

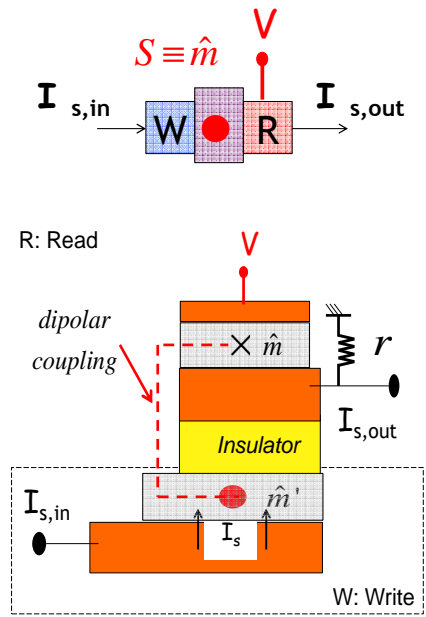

*Fig.2.5.2 All spin logic (ASL): Spin switch with spin voltages and spin currents as input and output.*

Spin currents could in principle carry more information than ordinary currents and enable devices one step closer to quantum information processing. However, unlike ordinary currents, spin currents die out within a spin coherence length which can vary widely from tens of nanometers to tens of microns depending on the material and the temperature of operation. As it stands, this could still allow information transfer up-to the first layer of interconnects (Metal layer 1, M1), but not for longer lengths (M2 and higher).

* datta@purdue.edu          ** Behtash.Behin-Aein@globalfoundries.com

**2.6 What do alternative switches have to offer?**

We have tried to present a general perspective on how *Write* and *Read* units can be integrated into switches with gain. But what do these alternative switches have to offer relative to the standard CMOS switches that are widely used?

***Energy-delay product:*** We started this chapter pointing out that a key roadblock on the path of miniaturization is the energy it takes to operate a switch. It is well-known that the switching energy can be reduced by going slow, so that the energy per se is not a fundamental property of a particular switch. It makes more sense to look at the energy-delay product. Consider for example [22] the charging of a capacitor $C$ through a resistance $R$ from a voltage $V$, for which it is well known that

$$\text{Energy}, E \sim QV$$
$$\text{Delay}, \tau \sim RC = RQ/V$$

Combining the two relations we obtain
$$\text{Energy-Delay Product:} \quad E\tau \sim Q^2 R \qquad (2.6.1)$$

suggesting that the energy-delay product is determined simply from two quantities:
    (1) How much charge $Q$ is being switched?
and   (2) What is the resistance $R$ through which it is being switched?

For CMOS switches the resistance $R$ is ~ tens of kilo-ohms, while the charge $Q$ is more difficult to estimate. In our introduction we used system level numbers to estimate the quantity $QV_{DD}$ as ~ 3000 eV, suggesting that $Q$ ~ 3000 electrons, since $V_{DD}$ ~ 1 volt. On the other hand if we look at the gate charge on an individual transistor it would be over an order of magnitude smaller. We believe the discrepancy is because the former estimate $Q$ includes additional parasitic charges.

How does this compare with spin switches? Spin switches being all metallic structures usually have lower resistances of several tens of ohms. But the charge $Q$ is ordinarily much larger, making $Q^2 R$ much larger. It has been shown [23] that the **minimum** charge $Q$ needed to switch a magnet through an ordinary spin-torque mechanism (Slonczewski spin transfer torque) is given by:
$$Q \geq 2qN_S \qquad (2.6.2)$$

where $N_S$ is the number of spins comprising the magnet which is related to the saturation magnetization through the relation (µ$_B$: Bohr magneton, $\Omega$:Volume of magnet)

\* *datta@purdue.edu*                     \*\* *Behtash.Behin-Aein@globalfoundries.com*

$$N_S = \frac{M_S \Omega}{\mu_B} \qquad (2.6.3)$$

Typical values of $M_s \sim 10^6$ A/m give an $N_s$ of about 100 spins in a volume of 1 nm$^3$. This means that even a magnet as small as $\sim$ 10 nm $\times$ 10 nm $\times$ 1 nm has $10^4$ spins, and from the inequality in Eq.(2.6.2), the charge $Q$ is at least 20,000 electrons well in excess of the CMOS numbers.

How fundamental is the inequality in Eq.(2.6.2)? One could understand this inequality by noting that the process of switching a magnet with a stream of incident electrons can be written as

$$\text{Incident Electrons} \; + \; N_s\,\hat{m} \; \rightarrow \; \text{Reflected Electrons} \; - \; N_s\,\hat{m}$$

During switching the magnet spin changes by $2N_s$ and it takes at least $2N_s$ electrons to conserve spin.

This argument, however, assumes that there is no other source of spin and the phenomenon of GSHE allows us to bypass this argument since the strong spin-orbit coupling provides a source of spin.

An electron on its way through the GSHE material gets deflected towards the magnet, flips its spin, has its spin randomized and then is deflected again by the spin-orbit 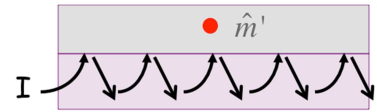 coupling towards the magnet and so on. In other words the same electron on its way through the GSHE material is incident repeatedly on the magnet and transfers many units of spin to it.

Indeed there is experimental evidence [14] that the GSHE material allows us to switch a magnet with less number of electrons than $2N_s$. It would seem that we could reduce the charge transferred from Eq.(2.6.2) to

$$Q \geq \frac{2qN_S}{\beta} \qquad (2.6.4)$$

\* *datta@purdue.edu*          \*\* *Behtash.Behin-Aein@globalfoundries.com*

However, the GSHE gain $\beta$ depends on the length of the magnet (Eq.(2.4.1)) which also makes the magnet longer and increases $N_S$. The possible improvement is thus useful but not unlimited. It thus seems that magnet-based switches will not provide the low power solution we are looking for unless more suitable *Write* mechanisms can be identified. Moreover, the dual MTJ *Read* unit fails with respect to another impressive characteristic of the CMOS switch: negligible standby power. These are the issues that need increased attention in the coming years.

***Beyond Boolean logic:*** Barring a major improvement is it worth pursuing alternative switches? We believe the answer is yes because of many other non-conventional applications that may be possible.

Consider for example the reconfigurable correlator shown in Fig.2.6.1 which should provide an output that correlates the incoming signal $\{X_n\}$ with a reconfigurable reference signal $\{Y_n\}$ stored in the $m_z$ of the switches that could be any string of +1's and -1's of length $N$, $N$ being a large number.

Since the output current of each Read unit is a product of $V(\sim X_n)$ and $m_z(\sim Y_n)$ it is determined by $X_n Y_n$ which are all added up to drive the output magnet. If the sequence $\{X\}$ is an exact match to $\{Y\}$, then the output voltage will be $N$, since every $X_n Y_n$ will equal +1, being either (+1)*(+1) or (-1)*(-1). If $\{X\}$ matches $\{Y\}$, in $(N-n)$, instances with $n$ mismatches, the output will be $N-2n$ since every mismatch lowers output by 2. If we set the threshold for the output magnet to $N-2N_e$ then the output will respond for all $\{X\}$ that matches the reference $\{Y\}$ within a tolerance of $N_e$ errors. The inset in Fig.2.6.1 shows an example with $N_e = 0$.

* *datta@purdue.edu*                    ** *Behtash.Behin-Aein@globalfoundries.com*

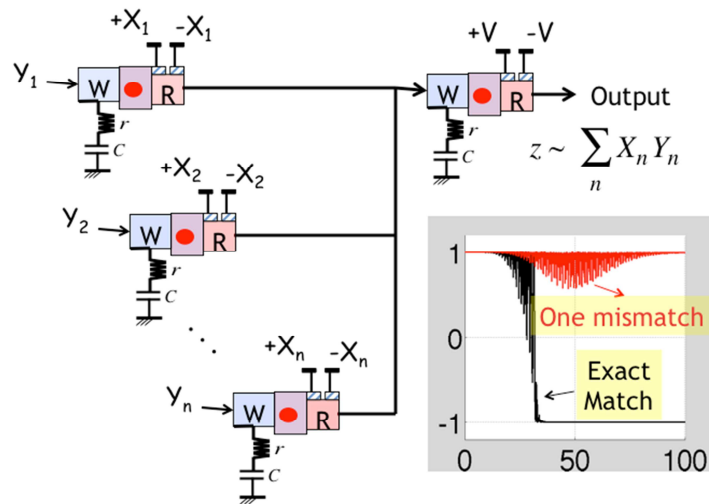

*Fig.2.6.1 An example of a device that could be implemented by interconnecting spin switches (Fig.2.4.2a) which should provide an output that correlates the incoming signal {$X_n$} with a reconfigurable reference signal {$Y_n$} stored in the switches. Inset shows response of output magnetization as a function of time (normalized). Threshold is adjusted such that the magnet is switched only if all 20 bits of {Y} match all 20 bits of {X}. With even one mismatch the output fails to switch. Note that no middle circuitry for signal conversion or amplification is involved. Reprinted with permission from*[5].*Copyright 2012, AIP Publishing LLC.*

This is a rather unique device which would allow us to correlate an input analog signal with a stored digital code to produce an analog output. This could be useful in mobile phones for decoding CDMA signals. Moreover it has an intriguing similarity to biological systems which correlate weak analog signals from cellular processes with a digital code stored in DNA molecules.

**Perspective**

But let us not go too far out on a limb with speculations. The objective here is simply to present our perspective viewing switches as *Write – Read* units underlining the key role played by gain and directionality in enabling large scale circuits. Hopefully this will help guide the search for new *Write* and *Read* mechanisms that could lead to fast low energy switches allowing miniaturization to continue for many more generations. But even otherwise, additional features like reconfigurability and non-volatility could enable new functionalities currently not available.

\* *datta@purdue.edu*                    \*\* *Behtash.Behin-Aein@globalfoundries.com*

**Summary**

This chapter presents our perspective on how *W* and *R* devices in general, spintronic or otherwise, can be integrated into switches having gain and directionality like transistors. Such switches could be interconnected to build complex circuits without external amplifiers or clocks. We start with a very brief and oversimplified discussion about CMOS transistors and argue that a CMOS switch can be viewed as an integrated *W-R* unit having an input-output asymmetry that give it gain and directionality. Next we discuss the standard *W* and *R* units used for magnetic memory devices and present one way to integrate them into a single unit with the input electrically isolated from the output. But this integrated *W-R* unit would not provide the key property of gain. We then show that the recently discovered giant spin Hall effect (GSHE) could be used to construct a *W-R* unit with gain and suggest other possibilities for spin switches with gain.

We end with a brief evaluation of these alternative switches in terms of possible applications. For conventional Boolean logic, at the present magnet-based switches will not provide the low power solution over standard CMOS switches unless more suitable *Write* mechanisms can be identified. On the other hand the non-volatility and reconfigurability of switches based on magnets is a novel feature that could enable a new class of circuits that are very different from those currently possible.

**Acknowledgements**

It is a pleasure to acknowledge a number of colleagues who have contributed to this work in different ways over many years: Sayeef Salahuddin, Angik Sarkar, Srikant Srinivasan, and Deepanjan Datta. S.D. and V.Q.D. are grateful for support from the Institute for Nanoelectronics Discovery and Exploration (INDEX) and the NSF-sponsored Center for Science of Information.

\* *datta@purdue.edu*                          \*\* *Behtash.Behin-Aein@globalfoundries.com*

\* *datta@purdue.edu*  \*\* *Behtash.Behin-Aein@globalfoundries.com*

\* *datta@purdue.edu*          \*\* *Behtash.Behin-Aein@globalfoundries.com*

**List of acronyms**

\*  *datta@purdue.edu*          \*\*  **Behtash.Behin-Aein@globalfoundries.com**